# Physical modeling of real-world slingshots for accurate speed predictions

Bob Yeats

(retired physicist, byeats@yahoo.com)

**Abstract**

We discuss the physics and modeling of latex-rubber slingshots. The goal is to get accurate speed predictions inspite of the significant real world difficulties of force drift, force hysteresis, rubber ageing, and the very non-linear, non-ideal, force vs. pull distance curves of slingshot rubber bands. Slingshots are known to shoot faster under some circumstances when the bands are tapered rather than having constant width and stiffness. We give both qualitative understanding and numerical predictions of this effect. We consider two models. The first is based on conservation of energy and is easier to implement, but cannot determine the speeds along the rubber bands without making assumptions. The second, treats the bands as a series of mass points subject to being pulled by immediately adjacent mass points according to how much the rubber has been stretched on the two adjacent sides. This is a classic many-body F=ma problem but convergence requires using a particular numerical technique. It gives accurate predictions (average error +2%, standard deviation 3%). The basis for these models involves measurements of pull-force vs. pull-distance. We will show how such measurements can be scaled for increased width or length, so that minimal calibration measurements are needed to support the models. With a good physical model of a slingshot one is easily able to test a wide range of parameters to determine how to get the fastest shots for a given pull force and draw length. We include examples of such calculations for popular slingshot rubber bands and tubes.

## Introduction

You can google "rubber slingshot physics" (or use google scholar) hoping to get technical discussions by physicists or engineers, but the only 'hits' are usually rather untechnical and too over-simplified to be helpful. Menzel in his website [1] gives a good general background on slingshots and includes a few plots of performance, but remarkably, it appears that serious slingshot analysis has escaped the scientific literature. So with this work, this retired solid-state physicist hopes to contribute a guide for navigating the complex world of real-world slingshots. My background includes designing and building novel semiconductor devices [2], temperature modeling of integrated circuit chips [3], and the development of new techniques for assessing the reliability of thin-film capacitors [4], for assessing the reliability of power transistors subjected to high electric fields [5-6], and for determination of the infant failure rate of large circuits at the level of 0.01% failure probability per transistor [7].

The interest in slingshots has increased in the last several years, I think mostly because of the demonstration of extreme pistol-like power by Joerg Sprave, who has both a blog and youtube channel [8]. There is also a slingshot forum [9] with a large following that focuses more on artistic slingshot creation, accurate shooting, and advice. Finally, there are many slingshot videos on youtube, including lighting a match with one shot and then putting it out with the next shot [10].

The speed of slingshot projectiles is of particular interest. Target shooters like flatter trajectories (and louder noises from their tin can targets), and hunters want more stopping power for small game. We consider two models for predicting slingshot speeds. The first is a simple energy conservation model (which we will call the "Energy model") in which the energy stored by the stretched rubber bands gets converted to the kinetic energy of the projectile, pouch, and rubber bands. This model cannot address the speed of the rubber bands at different positions along the rubber. Typical masses of the rubber bands used are similar to the masses used for projectiles, so one must try to take into account the kinetic energy of the rubber bands. We assume that when the projectile is released from the pouch near the end of the shot, that the speed along the band is proportional to the distance along the band to the fork. It turns out that the speed distribution along the rubber bands plays a significant role in allowing tapered slingshot bands to shoot faster, under some circumstances, than untapered bands. ("Tapered" bands are *flatbands* that are wider and stiffer near the fork than the pouch; the analog for *tube* bands are "pseudo-tapered" bands, which, for example, on each side of the slingshot have two tubes next to each other near the fork, and connect to only a single tube tied to the pouch).

The second model (which we will call the "F=ma model") *is* able to calculate the speed along the rubber bands, which is important for correctly calculating the speeds of tapered or pseudo-tapered slingshots. It considers the rubber bands to be made of a series of point masses. A given mass is pulled one way by one neighboring mass and the opposite way by the other neighboring mass. By calculating how much stretch there is between the neighboring adjacent two masses, one can determine the net force acting on the mass. This is an incremental time-evolving F=ma many-body problem. It turns out to be straight forward to set up the calculation for this model, but to get convergence requires a particular numerical technique (the "leapfrog" method [11]).

In the main body of this work we will first describe results and draw some conclusions. Toward the end we will give more details for successfully implementing the F=ma model.

## Discussion

### Force vs. Draw length curves

First some information about the kinds of rubber used in this work: We use latex rubber "flatbands" and "tubes". Flatbands are cut from sheets of latex, usually with a rotary cutter. The flatband rubber was made by Hygenic Corp. and purchased through simple-shot.com. There are two types: latex.03 (nominally "30 mil" thick, weighing 0.4439 g/inch$^2$), and latex.04 (nominal "40 mil" thick, weighing 0.5877 g/inch$^2$). The latex rubber tubes were purchased through Dankung.com and came from an unspecified vendor in Thailand. The various types of tubes we used are 2040 (density 0.2336 g/inch), 1842 (0.2710 g/inch), 3050 (0.3286 g/inch), 1745 (0.3385 g/inch), 2050 (0.4049 g/inch), and 4070 (0.5187 g/inch). All these densities were measured by me for my most recent batches. Typical variation between batches is around 1% – 2%, for both flatbands and tubes.

Figure 1 jumps right to the hysteresis effects that contribute to making slingshot modeling difficult. In this figure, there are 7 separate curves of pull force vs. "stretch factor" [SF].

(SF is the ratio of stretched to unstretched length. It is an important simplifying concept since it allows a universal curve to be used rather than requiring separate curves for each different unstretched length of bands.)

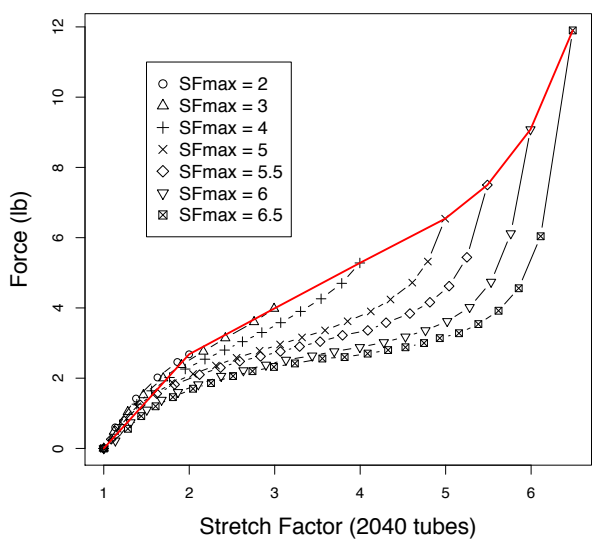

**Fig. 1.** Force vs. Stretch Factor for forward draw (top red curve) and reverse draws (black lower curves).

Each curve in Fig. 1 is associated with a maximum stretch factor. The first curve on the left results from pulling the band back in a single step to the maximum SF (=2 in this case). The "reverse draw" Force vs. Stretch Factor points are then acquired by starting with the full SF=2 draw and working back to the unstretched draw length (SF=1). The remaining curves are acquired in a similar way (always starting from zero force, with ~10 s delay, and jumping in a single step to the next maximum SF of a curve, then descending toward SF=1). The top red solid line curve connects the maximum SFs of the 7 curves. This curve can be thought of as the "forward draw" Force vs. SF curve (i.e., low to high force application order). The area under a force vs. SF curve, when multiplied by the unstretched band length, gives the energy stored in the rubber bands. This area is distinctly less for the reverse draw curves in Fig. 1 than for the forward draw curve, so any predictions of projectile speed based on the forward draw curve will significantly over-estimate the speed. I previously have described online some *forward* draw speed predictions [12], but did not have any measured speeds for comparison. To get the best estimate of stored energy for the *reverse* draw curves, one needs to use the curve corresponding to the maximum stretch factor used for the actual planned draw force (and SFmax). What causes the rapid reduction in force near SFmax for the reverse draw curves? Slingshot rubber is made from long polymer chains that are tangled together, and can be stretched and then want to retract. I picture a rubber band as being made of thousands of "micro-rubber-bands", that have velcro-like sides that let them stick to each other. When a rubber band is pulled back hard and the draw held for a short time, some of these micro-rubber-bands loose their grip on their neighbors and "fire", like mini slingshots. This energy will be lost as heat to the rubber band and the incremental contribution to the total reverse force will be lost. The best slingshot performance (e.g., greatest speed for a given pull force and draw length) usually occurs for SF > 5, but the bands also wear out faster at high SFs. Often SFmax in the 4.5– 5.5 range is chosen for a good compromise between higher speeds and faster wearout.

It is not trivial to generate a reverse draw curve, because when one holds a particular draw distance, the force required decreases rapidly, with a linear dependence on the logarithm of hold time (Fig. 2). The logarithmic time dependence means that the first few seconds have rapid change, but the final few seconds have much slower change. To hope to predict speeds, one must plan ahead to the hold time to be used for shooting. We choose 3.6 seconds (which is the first time point in Fig. 2). Our force meter takes 1-2 s to settle to a force value, so it would not be practical to measure a force at less than ~ 2 s . To help gather reverse draw data as quickly as possible, we take a movie for monitoring the amount of draw distance and force, and use a metronome to guide the timing. The movie camera, a force meter [13], and a position pointer are mounted to a sled that slides along a 4 foot ruler. The rubber band is connected to the force meter (usually through a pouch) and the movie shows the displayed force as well as the position, Fig. 3. Each reverse draw curve is held at SFmax for 36 s (= 30 ticks of a metronome at 1.2 s/tick), and then the reverse draw is started with 1 – 2 inch increments per step, and 2.4 seconds (= 2 ticks) per step. The 36 s initial hold slows down the drift so that there is not much drift subsequently during acquisition of the rest of the reverse draw curve. By calculating the ratio of the force at 36 s to the force at 3.6 s, the forces can be scaled back up by this ratio to what they would be expected to be at 3.6 s. The projected 3.6 s forces are what are shown in Fig. 1.

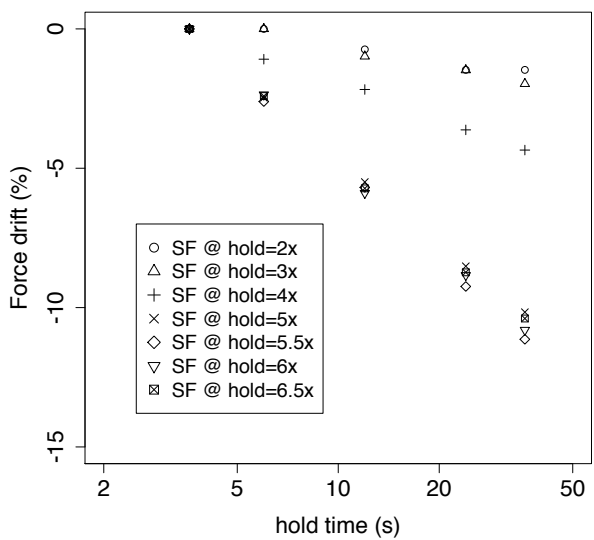

**Fig. 2.** Force drift vs. hold time at different stretch factors, for 2040 tubes.

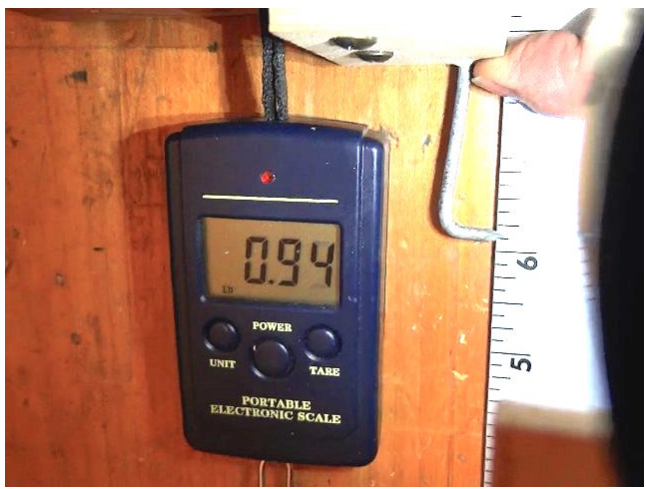

**Fig. 3.** Frame of movie for extracting force vs. draw length.

**Scaling of Force vs. SF curves**

There are 2 categories of bands used in slingshots: "flatbands" and "tubes", as described above. If a flatband doubles in width, the F vs. SF curves of Fig. 1 would change by the force doubling at each SF. For flatbands, the width can easily be doubled by cutting out a strip twice as wide from a sheet of latex (using a rotary cutter). For tubes, the width is doubled by using two bands in place of one, and again the force vs. SF curves of Fig. 1 would change by the force doubling at each SF. An alternative view, is that the force applied to a single tube splits into half that force applied to two identical tubes connected to it, so each of the doubled tubes has a SF associated with the half-force. If the *length* of a band doubles, Fig. 1 would remain unchanged. The same force is applied all the way along the double length band so each lengthwise half of the band would stretch by the same amount as the undoubled length band stretched. The result would be twice the length, twice the stretch, and unchanged stretch factor.

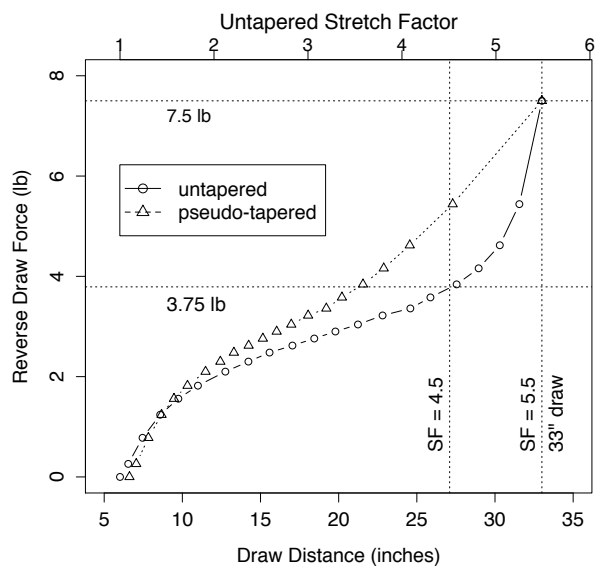

**Fig. 4.** Construction of pseudo-tapered force vs. draw distance curve from untapered curve, for 2040 tubes.

One does not need to take new measurements (beyond those in Fig. 1) to accommodate tapered or pseudo-tapered bands if one uses the scaling properties described here. For example, consider Fig. 4. Fig. 4 compares the Force vs. draw distance curves of 2040 tubes for pseudo-tapered and untapered bands each having the same pull force (7.5 lb) and draw distance *Xdraw* (from the fork). The pseudo-tapered curve is derived from the untapered curve. For the pseudo-tapered case, on each side of the slingshot a single band near the pouch (the "B=section") connects to two bands that connect to the fork (the "A-section"). Here we will take both the "A" section (attaching to the fork) and the "B" section (attaching to the pouch) to have the same unstretched length. A 7.5 lb force applied to the B-section splits equally into two 3.75 lb forces in the A-section. From the untapered curve in Fig. 4, we see that the A-section (with 2 bands, each having half the force) have a reduced

SF of SFa=4.5 compared to the untapered and B-section band SF of SFb=5.5. The required unstretched length, L0 = La0 + Lb0, can now be calculated from

$$La0*SFa + Lb0*SFb = Xdraw \quad [1]$$

With La0=Lb0 (for the unstretched lengths of the A and B sections), and Xdraw = 33, one finds La0=Lb0=3.3 inches for the pseudo-tapered case, and for the untapered case L0 = Xdraw/SFb = 33/5.5 = 6.0 inches. The remaining points for the two curves are found by choosing reduced force values, Fi, (Fi < 7.5 lb) and finding at which SFs Fi and Fi/2 intersect the untapered curve in Fig. 4. Since La0 and Lb0, and L0, are already known from the first full-draw step, multiplying any of these by the corresponding SF will give the stretched lengths, and with addition, the corresponding pull distance from the fork (< Xdraw).

## Why tapering may help speeds

Fig. 4 shows that the area under the pseudo-tapered curve is significantly larger than the area under the untapered curve. Since these areas equal the energy stored by the bands, in this case the pseudo-tapered bands store more energy for the same Xdraw and pull force. This allows the pseudo-tapered bands to *possibly* shoot faster than the untapered bands, provided the pseudo-tapered bands do not weigh too much more than the untapered bands. Fig. 5 shows similar curves, but for a reduced draw force of 5 lb (SFmax = 4) instead of 7.5 lb (SFmax = 5.5). Note that in this case it is the **un**tapered bands that have the larger area and store more energy. Fig. 6 gives a qualitative understanding of these results. At the right side of Fig. 6 near SF = 6, the force rises very steeply. We have exaggerated this steepness by adding the vertical dashed line between points A and B. There is no added area under the curve between points A and B, so no additional energy storage, yet the force required has risen by a few pounds. Hence, steep sections of the force vs. pull distance curve waste force and do not add much energy storage. Pseudo-tapering *reduces the steepness* of the Force vs. Draw Distance curve in steep regions (Fig. 4 near SF=5.5), but not so much for less steep regions, such as Fig. 5 near SFmax=4.

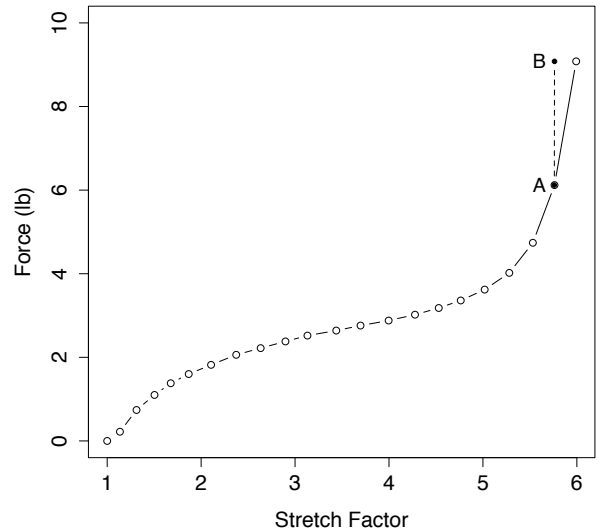

**Fig. 6.** Steep parts of the force vs. draw curve can have high forces but little area under the curve, so little energy storage.

There is another effect that favors tapering. Fig. 7 shows the F=ma model calculation of the speed distribution along the rubber bands when the bands reach their original unstretched length at the end of a shot. Fig. 7 shows that the pseudo-tapered case has very low speeds within the A-section bands, and even part way into the B-section bands. This has the effect of not wasting kinetic energy in the rubber material— as if the bands have much reduced mass. The energy saved then can go partly to extra speed in the projectile. Menzel [1] anticipated this effect and its implication that tapering would be more effective for *light weight* projectiles due to the increased importance of the rubber band mass. In Fig. 7, the tube type is 2040, the draw is 35 inches, the pull force is 9 lb, SFmax is 5.6, and in the pseudo-tapered section La0/L0 = 0.40, with L0=7.5 inches. The A-bands weigh 2.8 g, the B-bands weigh 2.1 g, and the 0.38 inch diameter steel ball projectile weighs 3.6 g, which is less than the total weight of the rubber bands. The elastic energy stored in the (reverse draw) rubber, is converted to the kinetic energy in the bands, the pouch, and the projectile, with the result that the untapered case has maximum projectile speed = 189 ft/s, compared to 216 ft/s for the pseudo-tapered case (a 14% increase). For a long meandering discussion with mostly favorable perceptions of tapering see the online forum topic in [14].

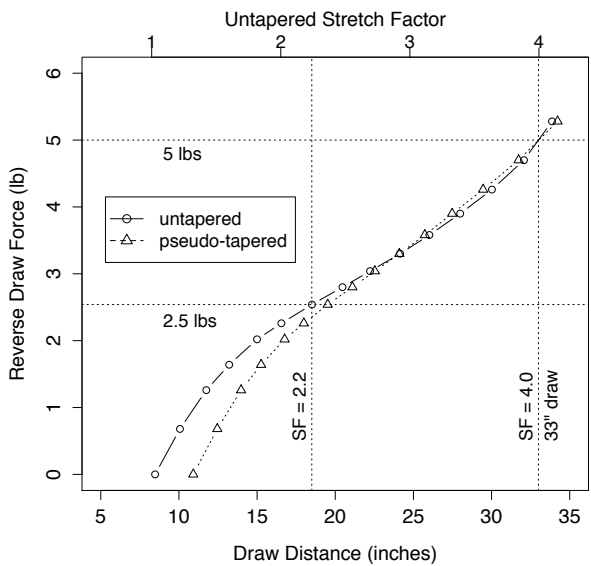

**Fig. 5.** Showing that at lower stretch factors, **un**tapered bands store more energy than pseudo-tapered bands

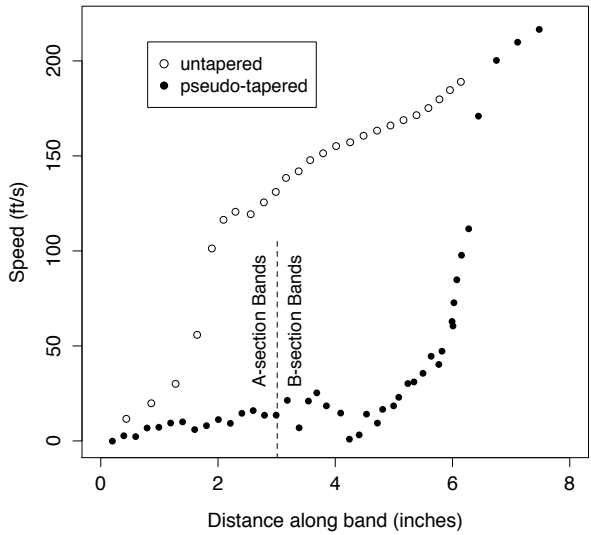

**Fig. 7.** For this case, the pseudo-tapered bands have very low speeds in the A-section of the bands, and partly into the B-section, thus energy is not wasted on the band kinetic energy

## Comparison of measured and calculated speeds

Fig. 8 compares the calculated and measured speeds using the F=ma model. There is good agreement over nearly a 2:1 range of speeds. Fig. 9 gives another view of this comparison, as it shows the distribution of the calculated to measured speed ratio, Vcalc/Vmeas.

The data underlying Figs. 8-9 come from Table 1 (for tubes) and Table 2 (for flatbands). Data includes results for tapered, pseudo-tapered, and untapered bands, and there are also a few cases in which (stainless steel) projectile ball diameters (Dball) are varied (with good predictions). The distribution in Fig. 9 is centered about 2% above the measured speeds and has a standard deviation of 3%. A plot similar to Fig. 9, but for the Energy model, is shown in Fig.

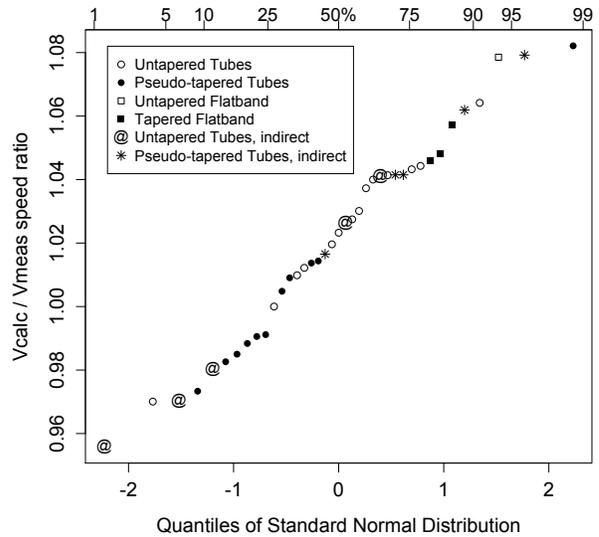

**Fig. 9.** Distribution of calculated to measured speed ratio for the F=ma speed model (median is 1.02, standard deviation is 3%).

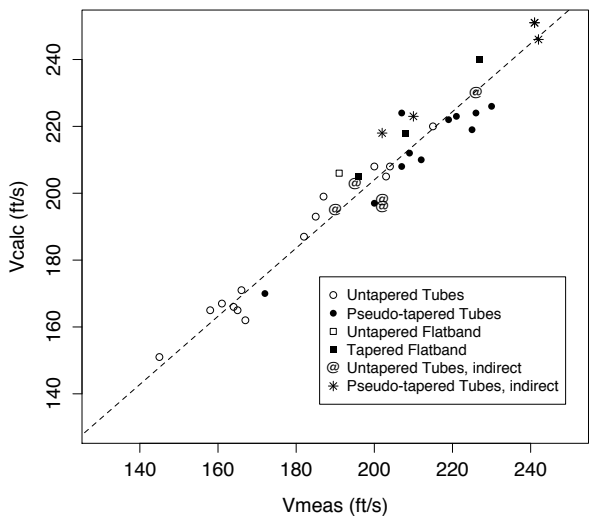

**Fig. 8.** Calculated vs. Measured projectile speeds for the F=ma model. Fitting line is drawn at Vcalc = 1.02 * Vmeas.

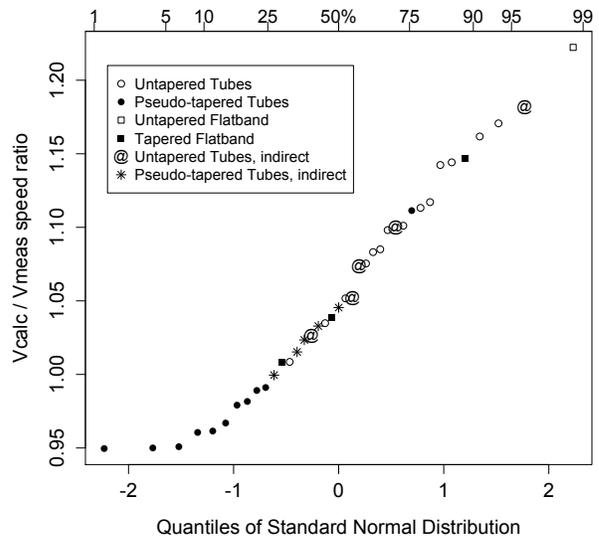

**Fig. 10.** Distribution of calculated to measured speed ratio for the Energy speed model (median is 1.05, standard deviation is 6%).

10, which shows the energy model distribution is centered about 5% above measured speeds, and has a standard deviation of about 6%. So the energy model is less accurate and has more spread than the F=ma model— a result that can be understood from the previous discussion of tapering, and also how the energy model needs to *assume* how speeds vary along the rubber. Once the F=ma model is implemented, there is no need to continue supporting the energy model. Calculation speeds are about 0.6 s per case, using the *R* programming language on a 2009 2.26 GHz MacBook Pro.

There is one category of data in Figs. 8-10 that needs explanation. Some of the data points are labeled "indirect". For these data points, the rubber type (e.g., 1842, 3050, and 4070) never had direct measurements of force vs. SF. Instead, their F vs. SF curves were inferred from measured density ratios of the rubber (e.g., g/inch of band length) compared to other rubbers that had direct F vs. SF measurements. The assumption is that the force at a given SF for similar rubbers is proportional to the density. Such a correction might also be used to correct for batch to batch variation (typically 1%-2%) for the same nominal type of rubber. It is interesting that these indirect cases have rather accurate speed predictions.

There is a very important contributor to the accuracy of our speed predictions. It is the "Force Adjustment Factor" (FAF), and is shown in Tables 1-2. The force vs. SF curves (e.g., Fig. 1) drift somewhat depending on recent history (hysteresis) and how much wearout has occurred. For example, while stretching a band to SF=5 does not usually change the unstretched length, stretching it to SF=6 could change the unstretched length by a few percent, that may take minutes, or sometimes hours, to return to its original unstretched length. Similarly, after a few hundred shots, the force at a given SF is usually reduced significantly (e.g., 10%-20%), even though the unstretched length may be the same. When I perform speed measurements (using a "Caldwell Ballistic Precision Chronograph"), I measure the force several times before taking the shot and measuring the speed. It frequently happens that the measured force is not quite what is expected from previous F vs. SF measurements. The FAF is a single scaling factor used to bring the measured F vs. SF curve into agreement with the force measured just before a speed measurement. Note that this "calibration" is ***not*** based on any speed knowledge, just the force measurements. After taking a force measurement, it is a good practice to release the (empty) pouch and wait ~ 10 s before taking new force or speed measurements. It seems that a *fully **un**stretched* band recovers to a constant state more so than a *partially* unstretched band. Speed measurements are usually surprisingly tight, typically a spread of ± 2% from the median for 5 measurements. To get an accurate draw length for the projectile speed measurements, I work on a table top with distances from the fork anchor marked every 1/8 inch. The draw distance is the distance from the fork anchor to the pouch rubber band tie. The speed measurements use a hold time of 3.6 s (3 ticks of the metronome), as do the force vs. SF curves as in Fig. 1.

Using a FAF seems a little like using a fudge factor, which most people would like to avoid. Unfortunately, the hysteresis and variablity of rubber over time ***does*** change the rubber characteristics, so the speed that the rubber will shoot at will vary. At least the FAF allows an easy single force measurement to correct for rubber changes, and we have demonstrated in Fig. 9 the degree of success this approach can have. Note that it is not very practical to take force vs. SF data on the normal completed 2-sided slingshot. Such data is actually taken for a single band then scaled for 2-bands and other lengths; 2-sided draw data would require holding the pouch with 22 lbs of force for 1745 tubes at SF=6.5, which is hard to do. (And even so, hysteresis issues would remain.) Note that a 10% error in force leads to a 10% error in stored rubber band energy, and about a 5% error in projectile speed.

Temperature affects slingshot performance. A thermodynamic argument [15] suggests that force at a given SF (and hold time) is proportional to the Kelvin temperature. This seems plausible from the data I have taken. We always adjust for temperature effects in our models and measurements.

## Calculated speed comparisons for different rubber types and tapering

Table 3 shows the highest calculated projectile speeds for fixed draw length and maximum allowed pull force, for various types of rubber, both tapered and untapered flatbands, and pseudo-tapered and untapered tubes. Performance is calculated for 35 inch draw distance from the fork, at 27°C, with a 1.3 g pouch, for 0.38 diameter stainless steel balls each weighing 3.6 g. The draw hold time was assumed to be 3.6 s. Pseudo-tapered bands have 2 A-section bands and 1 B-section band on each side of the fork. Total band(s) unstretched length is L0tot. For pseudo-tapered bands, La0/L0tot was varied through values of 0, 0.15, 0.30, 0.40, 0.50, 0.60, 0.70, and 0.85, while for flatbands, taper factors varied from 0.4 to 1.0 by 0.1. Flatband width at the pouch (Wpouch) varied from 0.3 to 1.5 inches by 0.1 inch. Pull forces (2-sided) were varied from 7.0 to 15.0 lb by 0.1 lb. Calculated results were filtered to exclude SFmax > 5.5, which we considered to be too fragile. This exclusion sometimes led to latex.03 or latex.04 bands having higher speeds at lower pull forces than the maximum allowed pull shown in column 2 of Table 3. For example, latex.04 at ≤13 lbs has 228 ft/s speed at 13 lb where SFmax is 5.01, but has 250 ft/s speed at 10.7 lb where SFmax is 5.50. So at SFmax=5.5, the speed is distinctly higher and comes with distinctly less pull force, 10.7 lb vs. 13 lb. Table 3 shows that latex.04 wins the speed race at all the listed allowed pulls, and most of these winners occur at reduced force (because of the SFmax ≤ 5.5 constraint). Latex.03 is only slightly worse than latex.04, and 1745 is similar to latex.03 near 13lb and 15 lb.

2050 is slightly worse than 1745. 2040 violates the SFmax ≤ 5.5 constraint for pulls > 8.6 lb, but at 8.5 lb is far better than 2050 or 1745 (mostly because SFmax is close to 5.5, unlike for 2050 or 1745 at 8.5 lb). For best speed, each type of band has a favored pull force corresponding to the allowed SFmax. Sheet latex (e.g., latex.03 or latex.04) has an advantage over tubes for this since it can be fine-tuned by adjusting the width or taper to adjust the SF, while tubes can only change the SF in *steps*, since the tube widths can only be adjusted in discrete *steps* in going from (e.g.) 2040 to 2050, to 1745 etc.. Larger SFs lead to shorter, lighter weight bands which helps projectile speed. It is easy to get especially high SFs in flatbands because they can easily be cut to extreme tapers and small widths, yet they are fundamentally more fragile than tubes because they have edges, unlike tubes, so tear more easily. Flatbands can be tapered without incurring a speed penalty, unlike pseudo-tapered tubes, which have extra mass at the connection between the A and B-section tubes, along with a reduced effective draw length. The extra mass is easy to include in the F=ma model. Table 3 shows that 1745 pseudo-tapered tubes incur about a 2% projectile speed penalty from the extra connection mass and reduced effective draw length.

**F=ma model details**

In this section we will describe in greater detail some aspects of the F = ma model.

There is an issue about how to interpolate to create a curve between two of the curves in Fig 1. For example, if one chose to operate at a draw length corresponding to SFmax=4.5, one would want to interpolate between the SFmax=4 and SFmax=5 curves. However, directly using SF by itself does not work since the SFmax=4 curve, unlike the SFmax=5 curve, does not have SFs between 4 and 5. Instead, interpolate based on normalized SF (=SF/SFmax within a curve). There is actually a double interpolation: First to use the same common set of normalized stretch factors for both curves, then for *each* normalized SF value, interpolate the force between the 2 curves based on the desired SFmax value for the new curve relative to the SFmax values of the two neighboring curves. The validity of this approach can be checked by using this interpolation technique to generate a curve for SFmax=5.5 and comparing it to the actual measured curve for SFmax=5.5 (as shown in Fig. 1). The curves would fall on top of each other.

Now we will describe how the calculation proceeds for a pseudo-tapered tube case. On each side of the slingshot there is an A-section attached to the fork and a B-section attached to the pouch. The A-section has two identical tubes next to each other and these connect to a single tube in the B-section. Each section is divided into evenly spaced point masses, with a single mass point where the A and B sections connect, and a final mass point at the pouch. Each mass point should have a mass consistent with the mass of the nearby rubber, plus possibly augmented by the pouch and projectile masses. Each mass point is connected to its two neighbors by short segments of the A or B rubber tube(s), and each of these segments must have its own force vs. SF curve calculated. By keeping track of the SFs for each segment, one can determine the forces pulling (in opposite directions) on a given mass point by its two neighbors. The net force on a mass point will lead to acceleration, and the right bookkeeping involving acceleration, speed and position will allow the new positions of each mass point to be determined. The new positions, will lead to a new set of SFs and hence a new set of forces and accelerations, and the process then repeats.

This process begins with a determination of the unstretched section lengths needed to get a given draw length at a given pull force (e.g., Eq. 1). At this initial time, all masses have zero speed, and all forces between masses are balanced. At t=0, the first iteration starts when the pouch is released. The force on the pouch is no longer balanced for the last mass point (at the pouch), so that point will move. However, the remaining mass points have not moved yet because of the balanced forces. With the next time increment (or iteration), the mass point next to the pouch has unbalanced force and will also move. After a number of time increments, all mass points in the band will have unbalanced forces and will be moving. All time increments are equal, with a typical value used = 1 ms. To get convergence (i.e., same speed result for different starting values, such as the number of mass points or the time increment value), it is critical that the right numerical method be used. In particular, I could not get this model to work until I used the "leapfrog" method, which is very clearly described in [11]. The main idea behind this method is use of a speed value at the midpoint of time increments rather than at either end.

More generally, the A-section rubber type does not need to be the same as for the B-section, and the A-section might even have a single stiff band rather than 2 less stiff bands. For *tapered **flatbands***, there are in essence as many sections as mass points, and the segments connecting mass points need to have their force vs. SF curves adjusted for varying band width. Lastly, if there are too many mass points, calculations for speed within a rubber band can get noisy, so it is less common to use > 30 mass points in an A or B band section.

**Summary**

We have shown how to realistically model real-world slingshots made out of currently popular types of rubber. We have shown how to accommodate tapering, hysteresis and gradual wearout, and have demonstrated that the "F=ma" model predicts measured speeds on average about 2% high, with a projectile speed distribution standard deviation of 3%. The speed distribution *along* the rubber bands that this model is able to calculate, helps explain why tapered or pseudo-tapered bands can shoot faster at a given draw and pull force than untapered bands. The performance of different shapes (flatband or tube) and types of rubber was compared, with the result that sheet latex outperforms tubes. Latex.04 was a little better than latex.03. When tubes are in the sweet spot of their force range, they are only slightly worse than flatbands. Flatbands, on the other hand, unlike tubes, can generally be *designed* to be in their sweet spot of force range by adjusting their width or taper when cutting them out of the sheet latex. Our model allows projectile speed optimization by calculation, which is much easier than experimentation.

## References

1. Melchior Menzel, in http://www.slingshotshooting.de/techstuff/science/science.html.

2. R. Yeats, K. Von Dessonneck, "Detailed performance characteristics of hybrid InP-InGaAsP APD's", IEEE Electron Device Letters, **EDL 2-10**, 268-271 (1981).

3. Bob Yeats, "Inclusion of topside metal heat spreading in the determination of HBT temperatures by electrical and geometric methods", 1999 GaAs IC Symposium Digest, 59-62.

4. Bob Yeats, "Assessing the reliability of silicon nitride capacitors in a GaAs IC process", IEEE transactions on electron devices, **45-4**, 939-946 (1998).

5. Bob Yeats, "The extraction of projected operating lifetime from accelerated stressing of a single device", 1997 MANTECH Digest, 175-179.

6. Bob Yeats, "Method for accelerated determination of GaAs PHEMT power slump reliability", Reliability of Compound Semiconductors Digest, 3-20 (2007).

7. B. Yeats, P. Chandler, M. Culver, D. D'Avanzo, G. Essilfie, C. Hutchinson, D. Kuhn, T. Low, T. Shirley, S. Thomas, W. Whiteley, "Reliability of InGaP-emitter HBTs", 2000 GaAs MANTECH Conference Digest, 131-135.

8. Joerg Sprave: http://www.slingshotchannel.com

9. http://www.slingshotforum.com

10. Bill Hays in https://youtu.be/jVj3MEmI46w

11. Sandvik, A.W., Numerical solutions of classical equations of motion, in http://physics.bu.edu/~py502/lectures3/cmotion.pdf .

12. Bob Yeats: http://slingshotforum.com/topic/22870-force-wall-effects-in-slingshot-operation/

13. The force meter shown in Fig. 3 is sold by multiple vendors on amazon.com, but the manufacturer is not identified, neither on the meter nor in its instructions. Its calibration was verified.

14. http://slingshotforum.com/topic/13242-testing-chinese-tubes/

15. A. N. Gent, "Rubber Elasticity: Basic concepts and behavior", in *Science and technology of rubber*, 3rd edition, pp. 1-27, eds. J. E. Mark and B. Erman, Elsevier (2005).

**Table 1.** Measured and calculated speed comparison using F=ma model for different band *tube* types and pseudo-tapering (*na* is the number of tubes per side in the A-section).

| A-bands | na | B-bands | La0 (in) | Lb0 (in) | T (°C) | Xdraw (in) | Pull (lb) | FAF | Dball (in) | Vmeas (ft/s) | Vcalc (ft/s) | Vcalc/ Vmeas |
|---------|----|---------|----------|----------|--------|------------|-----------|------|------------|--------------|--------------|--------------|
| ~ | 0 | 2040 | 0 | 6.66 | 20.7 | 33.33 | 6.17 | 0.846 | 0.38 | 161 | 167 | 1.037 |
| ~ | 0 | 2040 | 0 | 6.66 | 20.9 | 33.33 | 6.02 | 0.821 | 0.38 | 165 | 165 | 1.000 |
| 2040 | 2 | 2040 | 5 | 3.28 | 19.5 | 38.33 | 9.93 | 0.974 | 0.38 | 221 | 223 | 1.009 |
| ~ | 0 | 1745 | 0 | 7.44 | 16.9 | 37.2 | 9.94 | 0.786 | 0.38 | 187 | 199 | 1.064 |
| ~ | 0 | 1745 | 0 | 7.44 | 15 | 37.2 | 9.94 | 0.786 | 0.38 | 185 | 193 | 1.043 |
| ~ | 0 | 1745 | 0 | 7.44 | 15 | 37.2 | 9.94 | 0.786 | 0.44 | 166 | 171 | 1.030 |
| ~ | 0 | 1745 | 0 | 7.44 | 15 | 37.2 | 9.94 | 0.786 | 0.50 | 145 | 151 | 1.041 |
| ~ | 0 | 2050 | 0 | 7.9 | 15.2 | 39.5 | 10.59 | 0.8 | 0.38 | 200 | 208 | 1.040 |
| ~ | 0 | 2050 | 0 | 7.9 | 15.2 | 39.5 | 10.59 | 0.8 | 0.44 | 182 | 187 | 1.027 |
| ~ | 0 | 2050 | 0 | 7.9 | 15.2 | 39.5 | 10.59 | 0.8 | 0.50 | 158 | 165 | 1.044 |
| ~ | 0 | 2040 | 0 | 6.75 | 20.4 | 33.75 | 5.78 | 0.793 | 0.38 | 167 | 162 | 0.970 |
| ~ | 0 | 2040 | 0 | 6.75 | 22.1 | 33.75 | 6.06 | 0.827 | 0.38 | 164 | 166 | 1.012 |
| 2040 | 2 | 2040 | 5 | 3.28 | 20.3 | 38.33 | 9.90 | 0.966 | 0.38 | 219 | 222 | 1.014 |
| 2050 | 1 | 2040 | 5.13 | 3.38 | 24.5 | 38.33 | 8.52 | 0.884 | 0.38 | 212 | 210 | 0.991 |
| ~ | 0 | 2050 | 0 | 7.9 | 24 | 39.5 | 11.89 | 0.871 | 0.38 | 215 | 220 | 1.023 |
| ~ | 0 | 1745 | 0 | 7.44 | 23.4 | 37.2 | 10.87 | 0.84 | 0.38 | 204 | 208 | 1.0200 |
| 2050 | 1 | 2040 | 5.65 | 2.47 | 24.8 | 37 | 9.15 | 0.865 | 0.38 | 209 | 212 | 1.014 |
| 1745 | 1 | 2040 | 5.03 | 2.17 | 24.7 | 37 | 10.32 | 0.86 | 0.38 | 230 | 226 | 0.983 |
| 2050 | 1 | 2040 | 5.65 | 2.47 | 24.3 | 37 | 8.76 | 0.83 | 0.38 | 207 | 208 | 1.005 |
| 1745 | 1 | 2040 | 4.94 | 2.16 | 23.9 | 37 | 10.17 | 0.834 | 0.38 | 226 | 224 | 0.991 |
| 1745 | 1 | 2040 | 4.94 | 2.16 | 23.9 | 37 | 10.17 | 0.834 | 0.44 | 200 | 197 | 0.985 |
| 1745 | 1 | 2040 | 4.94 | 2.16 | 23.9 | 37 | 10.17 | 0.834 | 0.50 | 172 | 170 | 0.988 |

| A-bands | na | B-bands | La0 (in) | Lb0 (in) | T (°C) | Xdraw (in) | Pull (lb) | FAF | Dball (in) | Vmeas (ft/s) | Vcalc (ft/s) | Vcalc/ Vmeas |
|---|---|---|---|---|---|---|---|---|---|---|---|---|
| 1745 | 1 | 2040 | 4.94 | 2.16 | 23.5 | 37 | 9.74 | 0.798 | 0.38 | 225 | 219 | 0.973 |
| ~ | 0 | 1842[a] | 0 | 7 | 21.7 | 35 | 8.47 | 0.997 | 0.38 | 190 | 195 | 1.026 |
| ~ | 0 | 3050[b] | 0 | 7 | 23.3 | 35 | 9.92 | 0.786 | 0.38 | 202 | 196 | 0.97 |
| 4070[c] | 1 | 1745 | 4.7 | 3.13 | 26.3 | 33.25 | 14.1 | 0.978 | 0.38 | 241 | 251 | 1.041 |
| 1745 | 1 | 1842[a] | 5.19 | 2.25 | 25.6 | 33.25 | 10.0 | 0.982 | 0.38 | 210 | 223 | 1.062 |
| ~ | 0 | 1842[a] | 0 | 7 | 30.1 | 34 | 8.89 | 1.055 | 0.38 | 202 | 198 | 0.980 |
| 1745 | 1 | 1842[a] | 3.88 | 3.88 | 29.3 | 34 | 9.39 | 1.024 | 0.38 | 202 | 218 | 1.079 |
| 4070[c] | 1 | 1745 | 4.7 | 3.13 | 26.3 | 34 | 14.1 | 0.978 | 0.38 | 241 | 251 | 1.041 |
| ~ | 0 | 4070[c] | 0 | 8.31 | 26.1 | 34 | 16.89 | 1.055 | 0.38 | 226 | 230 | 0.956 |
| 4070[c] | 1 | 1745 | 4.7 | 3.13 | 28.9 | 33.25 | 13.74 | 0.975 | 0.38 | 242 | 246 | 1.017 |
| ~ | 0 | 1745 | 0 | 7.44 | 28.4 | 34 | 10.53 | 0.885 | 0.38 | 203 | 205 | 1.010 |
| 2040 | 2 | 1745 | 3.75 | 3.75 | 27.2 | 33.25 | 11.63 | 0.963 | 0.38 | 207 | 224 | 1.082 |
| ~ | 0 | 1842[a] | 0 | 7 | 27 | 36 | 9.42 | 1.042 | 0.38 | 195 | 203 | 1.041 |

[a] 1824 tubes have density 1.16 times density of 2040. Speed calculations are indirect, being based on 2040 force measurements scaled by this density ratio.

[b] 3050 tubes have density 0.9737 times density of 1745. Speed calculations are indirect, being based on 1745 force measurements scaled by this density ratio.

[c] 4070 tubes have density 1.532 times density of 1745. Speed calculations are indirect, being based on 1745 force measurements scaled by this density ratio.

**Table 2.** Measured and calculated speed comparison using F=ma model for different band *flatband* types and tapering.

| Band Type | L0 (in) | Taper | Wpouch (in) | T (°C) | Xdraw (in) | Pull (lb) | FAF | Dball (in) | Vmeas (ft/s) | Vcalc (ft/s) | Vcalc/ Vmeas |
|---|---|---|---|---|---|---|---|---|---|---|---|
| latex.03 | 7.32 | 1 | 0.88 | 22.6 | 36.6 | 10.11 | 0.976 | 0.38 | 191 | 206 | 1.079 |
| latex.03 | 7.56 | 0.713 | 0.67 | 19.7 | 37.8 | 9.73 | 1.218 | 0.38 | 227 | 240 | 1.057 |
| latex.03 | 8.00 | 0.639 | 0.76 | 28.1 | 34.0 | 10.4 | 1.177 | 0.38 | 208 | 218 | 1.048 |
| latex.04 | 8.25 | 0.850 | 0.71 | 27.6 | 34.0 | 11.17 | 1.064 | 0.38 | 196 | 205 | 1.046 |

**Table 3.** Calculated performance comparison of different band types and tapers, restricted to SFmax < 5.5. Winning speed for each force is shown in **bold red**. Latex.04 is the winner. Most latex.04 winners even occur at reduced force to allow use of highest allowed SFmax (= 5.5); e.g. 10.7 lb instead of 13 lb. Latex.03 is only slightly worse than latex.04, and 1745 is similar to latex.03 near 13 lb and 15 lb. 2050 is slightly worse than 1745. 2040 violates the SFmax < 5.5 constraint for pulls > 8.6 lb, but at 8.5 lb is only slightly worse than latex.03 and is far better than 2050 or 1745 (mostly because SFmax is close to 5.5, unlike for 2050 or 1745 at 8.5 lb).

| | | Best v of untapered, tapered, or pseudo-tapered | | | | | | | Best *un*tapered v |
|---|---|---|---|---|---|---|---|---|---|
| Band Type | Allowed Pull (lb) | Actual Pull (lb) | SFmax | v (ft/s) | La0/L0tot | L0tot | Taper | Wpouch (in) | v (ft/s) |
| 2040 | ≤8.5 | 8.5 | 5.44 | 212 (208[a]) | 0.40 | 7.80 | ~ | ~ | 189 |
| 2050 | ≤8.5 | 8.5 | 3.26 | 160 | 0 | 10.75 | ~ | ~ | 160 |
| 2050 | ≤11 | 11.0 | 4.02 | 203 | 0 | 8.71 | ~ | ~ | 203 |
| 2050 | ≤13 | 13.0 | 4.72 | 229 (226[a]) | 0.15 | 7.95 | ~ | ~ | 224 |
| 2050 | ≤15 | 15.0 | 5.32 | 250 (245[a]) | 0.30 | 7.60 | ~ | ~ | 230 |
| 1745 | ≤8.5 | 8.5 | 3.40 | 176 | 0 | 10.28 | ~ | ~ | 176 |
| 1745 | ≤11 | 11.0 | 4.27 | 217 (214[a]) | 0.15 | 8.80 | ~ | ~ | 215 |
| 1745 | ≤13 | 13.0 | 4.98 | 239 (235[a]) | 0.30 | 8.16 | ~ | ~ | 223 |
| 1745 | ≤15 | 15.0 | 5.38 | 256 (250[a]) | 0.30 | 7.48 | ~ | ~ | 224 |
| latex.03 | ≤8.5 | 8.5 | 5.14 | 220 | ~ | 7.41 | 0.6 | 0.7 | 195 |
| latex.03 | ≤11 | 10.1<br>11.0 | 5.50<br>5.20 | 245<br>228 | ~ | 8.58<br>7.28 | 0.4<br>0.6 | 0.8<br>0.9 | 211 @ 11 lb |
| latex.03 | ≤13 | 12.6<br>13.0 | 5.48<br>4.89 | 246<br>221 | ~ | 8.64<br>7.37 | 0.4<br>0.8 | 1.0<br>1.1 | 223 @ 12.6 lb |
| latex.03 | ≤15 | 13.9<br>15.0 | 5.50<br>5.41 | 247<br>244 | ~ | 8.86<br>6.69 | 0.4<br>0.7 | 1.1<br>1.2 | 235 @ 13 lb |
| latex.04 | ≤8.5 | 8.5 | 5.25 | **227** | ~ | 7.19 | 0.6 | 0.5 | 196 @ 8.1 lb |
| latex.04 | ≤11 | 10.7<br>11.0 | 5.50<br>4.74 | **250**<br>214 | ~ | 7.12<br>7.65 | 0.5<br>0.8 | 0.6<br>0.7 | 211 @ 11 lb |
| latex.04 | ≤13 | 10.7<br>13.0 | 5.50<br>5.01 | **250**<br>228 | ~ | 7.12<br>7.18 | 0.5<br>0.8 | 0.6<br>0.8 | 225 @ 13 lb |
| latex.04 | ≤15 | 14.2<br>15.0 | 5.48<br>5.14 | **251**<br>236 | ~ | 8.58<br>6.97 | 0.4<br>0.8 | 0.8<br>0.9 | 232 @ 15 lb |

[a]Including realistic speed penalty at A-B section joint for pseudo-taper tubes caused by 0.6 g extra (2-sided) mass from 0.5 inch diameter leather donut coupler and extra tie-rubber at donut. Also includes 0.75 inch estimated reduction in effective draw length by donut. Flatbands do not have any tapering penalties, except faster wearout at high SFs.